\documentclass[english,preprint,aps]{revtex4}
\usepackage[T1]{fontenc}
\usepackage[english]{babel}
\usepackage{amsmath}
\usepackage{amssymb}
\usepackage{epsfig}

\makeatletter
\@ifundefined{textcolor}{}
{%
 \definecolor{BLACK}{gray}{0}
 \definecolor{WHITE}{gray}{1}
 \definecolor{RED}{rgb}{1,0,0}
 \definecolor{GREEN}{rgb}{0,1,0}
 \definecolor{BLUE}{rgb}{0,0,1}
 \definecolor{CYAN}{cmyk}{1,0,0,0}
 \definecolor{MAGENTA}{cmyk}{0,1,0,0}
 \definecolor{YELLOW}{cmyk}{0,0,1,0}
 }

\usepackage{amsfonts}\usepackage{dcolumn}\usepackage{bm}

\makeatother

\usepackage{babel}

\makeatother

\usepackage{babel}

\begin{document}

\title{Quasi Sturmian Basis in Two-Electron Continuum Problems}

\author{A.~S.~Zaytsev$^{1}$,  L. U.
Ancarani$^{2}$, and S.~A.~Zaytsev$^{1}$}

\affiliation{ $^{1}$Pacific National University, Khabarovsk, 680035, Russia}


\affiliation{$^{2}$Equipe TMS, SRSMC, UMR CNRS 7565, Universit\'{e}
de Lorraine, 57078 Metz, France}

\date{\today}
\begin{abstract}
A new type of basis functions is proposed to describe a two-electron
continuum which arises as a final state in electron-impact
ionization and double photoionization of atomic systems. We name
these functions, which are calculated in terms of the recently
introduced Quasi Sturmian functions, Convoluted Quasi Sturmian
functions (CQS). By construction, the CQS functions look
asymptotically like a six-dimensional spherical wave. The driven
equation describing an $(e, 3e)$ process on helium in the framework
of the Temkin-Poet model has been solved numerically using
expansions on the basis CQS functions. The convergence behavior of
the solution has been examined as the size of the basis has been
increased. The calculations show that the convergence rate is
significantly improved by introducing a phase factor corresponding
the electron-electron interaction into the basis functions. Such a
modification of the boundary conditions leads to appreciable change
in the magnitude of the solution.

\end{abstract}
\maketitle

\section{Introduction}
The Coulomb three-body scattering problem is one of the most
fundamental outstanding problems in theoretical atomic and molecular
physics. The primary difficulty in description of three charged
particles in the continuum is imposing appropriate asymptotic
behaviors of the wave function.

Several {\em ab initio} methods are developed for constructing
solutions to the three-body scattering problem (see the review
\cite{Bray:2012}). The exterior complex scaling (ECS) method (see
\cite{McCurdy:2004B} and references therein) allows the problem to
be solved without explicit use of the asymptotic boundary
conditions. Specifically, ECS recasts the original problem into a
boundary problem with zero boundary conditions. (For an extension of
ECS to the case of long-range Coulombic interactions see
\cite{Volkov:2009,Elander:2009}.) Some of the other methods use a
product of two fixed charge Coulomb waves to approximate the
asymptotic three-body continuum state. The convergent close coupling
(CCC) \cite{CCC:1992,CCC:2002,CCC:2004} and the Coulomb-Sturmian
separable expansion \cite{Papp:2001,Papp:2002} and the $J$-matrix
\cite{Zaytsev:2007,Mengoue:2011} methods treat the problem in the
Laguerre basis representation. The latter two methods transform the
original problem to a Lippmann-Schwinger-type integral equation
whose kernel seems to be generally non-compact. Alternatively, the
Generalized Sturmian Functions (GSF) method
\cite{Frapiccini:2010,Gasaneo:2013B} converts the problem into an
inhomogeneous Schr\"{o}dinger equation with a square integrable
driven term. One-particle generalized Sturmian functions with an
appropriate asymptotic behavior are obtained (numerically) as
eigensolutions of a Sturm-Liouville problem. The GSF method driven
equation is solved by an expansion into a basis set of two-particle
functions, which are products of two generalized Sturmian functions
that both satisfy outgoing-wave boundary conditions
\cite{Gasaneo:2013}.


In the present paper in order to describe a Coulomb three-body
system continuum we propose a set of two-particle functions, which
are calculated by using recently introduced so called Quasi Sturmian
(QS) functions \cite{DelPunta:2014}. The latter satisfy a two-body
inhomogeneous Schr\"{o}dinger equation with a Coulomb potential and
an outgoing-wave boundary condition. Specifically, the two-particle
basis functions are obtained, by analogy with the Green's function
of two non-interacting hydrogenic atomic systems, as a convolution
integral of two one-particle QS functions. The QS functions have the
merit that they are expressed in closed form, which allows us to
find an appropriate integration path that is useful for numerical
calculations of such an integral representation. We name these basis
functions Convoluted Quasi Sturmian (CQS) functions. Note that by
construction, the CQS function (unlike a simple product of two
one-particle ones) looks asymptotically (as the hyperradius $\rho
\rightarrow \infty$) like a six-dimensional outgoing spherical wave.

We apply these CQS functions to the solution of a problem of double
ionization of He in the framework of the Temkin-Poet model. We solve
the driven equation describing an $(e, 3e)$ process
\cite{Gasaneo:2013} by an expansion into the basis set of CQS
functions and explore the convergence properties of the expansion.
Note that the CQS functions asymptotic behavior in the so called
three-body region $\Omega_0$ where all three particles are well
separated is not correct since it misses out the phase factor,
corresponding to the Coulomb interelectronic interaction. Therefore,
the expansion method effectiveness is open to question. In order to
improve the convergence rate, we equip the basis functions with the
phase factor corresponding to the potential $\frac{1}{r_{12}}$.

The paper is arranged as follows. In Sec.~II we present the
three-body driven equations \cite{Gasaneo:2013} and \cite{Selles}
whose solutions possess all the information for the $(e, 3e)$
process on helium and that for the one-photon ionization,
respectively. Here we suggest the CQS functions which form a basis
set used for solving these equations. The CQS functions are expanded
in a series of products of the single-particle Laguerre basis
functions. A useful integral representation is also introduced for
the CQS functions. The asymptotic behavior of the basis functions in
the region $\Omega_0$ is deduced from their integral representation.
In this section we also propose a modification of the basis
functions which allows to take into account the $e-e$ interaction.
The use of both original and modified versions of the basis
functions in solving the $s$-wave driven equation
\cite{Gasaneo:2013} is considered in Sec~III. Finally, Sec.~IV
provides a summary. Atomic units are assumed throughout.

\section{Quasi Sturmian basis functions}

\subsection{Driven equations}
Electron-impact ionization and double photoionization of atomic
systems can be cast as an inhomogeneous three-body Schr\"{o}dinger
equation with a square integrable right hand side. For example, in
the approach \cite{Gasaneo:2013} to the (e, 3e) process on helium,
the four-body Schr\"{o}dinger equation is reduced to the following
driven equation for the three-body system $(e^-,e^-,{\mbox
He}^{++})=(1,2,3)$:
\begin{equation}\label{ME}
    \left[E- \hat{H} \right]\Phi^{(+)}({\bf r}_1, {\bf r}_2)
     = \hat{W}_{fi}({\bf r}_1, {\bf r}_2)\Phi^{(0)}({\bf r}_1, {\bf r}_2).
\end{equation}
$E=\frac{k_1^2}{2}+\frac{k_2^2}{2}$ is the energy of the two ejected
electrons. The three-body helium Hamiltonian is given by
\begin{equation}\label{H}
    \hat{H}=\hat{H}_1+\hat{H}_2+\frac{1}{r_{12}},
\end{equation}
\begin{equation}\label{H0}
    \hat{H}_j=-\frac{1}{2}\triangle_{r_j}-\frac{2}{r_j}, \quad j=1,
    2.
\end{equation}
$\Phi^{(0)}({\bf r}_1, {\bf r}_2)$ represents the ground state of
the helium atom. The perturbation operator $\hat{W}_{fi}$ is written
as
\begin{equation}\label{Wfi}
    \hat{W}_{fi}({\bf r}_1, {\bf
    r}_2)=\frac{1}{(2 \pi)^3}\frac{4 \pi}{q^2}(-2+e^{i{\bf q}\cdot {\bf r}_1}
     +e^{i{\bf q}\cdot {\bf r}_2}),
\end{equation}
where ${\bf q}={\bf k}_i-{\bf k}_f$ is the transferred momentum,
${\bf k}_i$ and ${\bf k}_f$ are the momenta of the incident and
scattered electrons.

In turn, the one-photon ionization problem also takes the form of
the driven equation \cite{Selles}
\begin{equation}\label{ME2}
    \left[E- \hat{H} \right]\Phi^{(+)}({\bf r}_1, {\bf r}_2)
     = \frac{1}{2}\vec{\mathcal{E}}_0 \cdot \vec{D}_G \Phi^{(0)}({\bf r}_1, {\bf r}_2),
\end{equation}
where $\vec{\mathcal{E}}_0$ is the amplitude of the electric-field
vector and $\vec{D}_G$ is the dipole operator.

\subsection{Convoluted Quasi Sturmians}

Our method of solving the driven equations (\ref{ME}) and
(\ref{ME2}) is to expand the solution
\begin{equation}\label{FE}
    \Phi^{(+)}({\bf r}_1, {\bf r}_2)=\sum \limits _{L, \ell, \lambda} \sum \limits_{n_1,
    n_2=0}^{N-1}
     C_{n_1 n_2}^{L(\ell_1 \ell_2)}\left|n_1 \ell_1 n_2 \ell_2; LM \right>_{Q}
\end{equation}
on the basis
\begin{equation}\label{QSANG}
    \left|n_1 \ell_1 n_2 \ell_2; LM \right>_{Q} \equiv \frac{Q_{n_1
    n_2}^{\ell_1
  \ell_2 (+)}(E; r_1, r_2)}{r_1 r_2}\mathcal{Y}^{\ell_1 \ell_2}_{LM}(\hat{\bf r}_1,
  \hat{\bf r}_2),
\end{equation}
\begin{equation}\label{SH}
\mathcal{Y}^{\ell_1 \ell_2}_{LM}(\hat{\bf r}_1,\hat{\bf r}_2)=\sum
\limits_{m_1 m_2=M}(\ell_1 m_1 \ell_2 m_2\left|LM \right.)Y_{\ell_1
m_1}(\hat{\bf r}_1)Y_{\ell_2 m_2}(\hat{\bf r}_2).
\end{equation}
Each function $Q_{n_1 n_2}^{\ell_1 \ell_2 (+)}$ is assumed to satisfy
the radial equation
\begin{equation}\label{PQSE}
    \left[E-\hat{h}_1^{\ell_1}-\hat{h}_2^{\ell_2} \right]Q_{n_1 n_2}^{\ell_1 \ell_2 (+)}(E; r_1,
    r_2)=\frac{\psi_{n_1}^{\ell_1}(r_1)\psi_{n_2}^{\ell_2}(r_2)}{r_1 r_2},
\end{equation}
where
\begin{equation}\label{hxy}
 \hat{h}^{\ell}=-\frac{1}{2}\frac{\partial^2}{\partial
 r^2}+\frac{1}{2}\frac{\ell(\ell+1)}{r^2}-\frac{2}{r},
\end{equation}
$\psi_{n}^{\ell}$ are the Laguerre basis functions ($b$ is a real scale parameter)
\begin{equation}\label{Lbf}
    \psi_{n}^{\ell}(r)=\left[(n+1)_{2\ell+1}
    \right]^{-\frac{1}{2}}(2br)^{\ell+1}e^{-b r}L_n^{2\ell+1}(2br),
\end{equation}
which are orthogonal with the weight $\frac{1}{r}$:
\begin{equation}\label{Orth}
    \int \limits _{0}^{\infty}dr
    \psi_{n}^{\ell}(r)\frac{1}{r}\psi_{m}^{\ell}(r)=\delta_{nm}.
\end{equation}

\begin{figure*}[ht]
\centerline{\psfig{figure=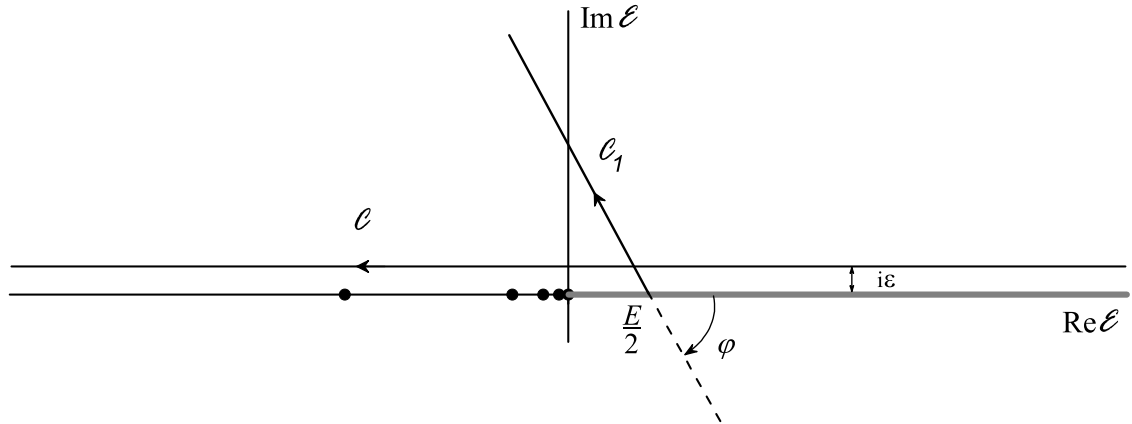,width=1\textwidth}}
\caption{$\mathcal{C}$ is the straight-line path of integration of
the convolution integral (\ref{GC}). The bound-state poles of
$\widehat{G}^{\ell_1(+)}(\sqrt{2\mathcal{E}})$ are depicted as full
circles. The grey line is the unitary branch cut. A part of the
rotated contour $\mathcal{C}_{1}$ (the dashed line) lies in the
region of unphysical energies.} \label{fig1}
\end{figure*}

In order to obtain the $Q_{n_1 n_2}^{\ell_1 \ell_2 (+)}$ with the
outgoing-wave boundary condition we use the Green's function
$\hat{G}^{\ell_1 \ell_2(+)}(E)$ which can be expressed in the form
of the convolution integral
\cite{Baz_Zeldovich_Perelomov,Shakeshaft:2004}
\begin{equation}\label{GC}
\hat{G}^{\ell_1 \ell_2(+)}(E)=\frac{1}{2 \pi i}\int \limits_
{\mathcal{C}}d\mathcal{E}\,\hat{G}^{\ell_1(+)}(\sqrt{2\mathcal{E}})
\hat{G}^{\ell_2(+)}(\sqrt{2(E-\mathcal{E})}),
\end{equation}
where the path of integration $\mathcal{C}$ in the complex energy
plane $\mathcal{E}$ runs slightly above the branch cut and
bound-states poles of $\hat{G}^{\ell_1(+)}$ (see Fig.~\ref{fig1}).
Applying the Green's function operator (which is the inverse of the
operator in the left-hand side of (\ref{PQSE})) onto both sides of
equation (\ref{PQSE}), we find that
\begin{equation}\label{QC}
Q^{\ell_1 \ell_2(+)}_{n_1 n_2}(E; r_1, r_2)=\frac{1}{2 \pi i}\int
\limits_
{\mathcal{C}}d\mathcal{E}\,Q^{\ell_1(+)}_{n_1}(\sqrt{2\mathcal{E}};
r_1)Q^{\ell_2(+)}_{n_2}(\sqrt{2(E-\mathcal{E})}; r_2),
\end{equation}
where the one-particle quasi Sturmian function $Q^{\ell(+)}_{n}$ is
defined by \cite{DelPunta:2014}
\begin{equation}\label{Qn}
    Q^{\ell(\pm)}_n(k, r)=\int \limits
    _{0}^{\infty}dr'\,G^{\ell(\pm)}(k;r,r')\frac{1}{r'}\psi_{n}^{\ell}(r').
\end{equation}
We name the basis functions (\ref{QC}) Convoluted Quasi Sturmian
(CQS).

\subsection{Laguerre basis expansion} CQS functions can be
expanded in terms of the Laguerre basis functions (\ref{Lbf}) as
\begin{equation}\label{CQSS}
    Q^{\ell_1 \ell_2(+)}_{n_1 n_2}(E; r_1, r_2)=\sum
    \limits_{m_1,m_2=0}\psi_{m_1}^{\ell_1}(r_1)\psi_{m_2}^{\ell_2}(r_2)
    G_{m_1 m_2, n_1 n_2}^{\ell_1 \ell_2(+)}(E).
\end{equation}
The coefficients $G_{m_1 m_2, n_1 n_2}^{\ell_1 \ell_2(+)}(E)$ are
the matrix elements of the Green's function (\ref{GC}) over the
functions
\begin{equation}\label{LLbf}
\frac{\psi_{n_1}^{\ell_1}(r_1)\psi_{n_2}^{\ell_2}(r_2)}{r_1 r_2}
\end{equation}
and can be calculated using the convolution
\cite{Papp:2002,Zaytsev:2007,Mengoue:2011}
\begin{equation}\label{GMECI}
    G_{m_1 m_2, n_1 n_2}^{\ell_1 \ell_2(+)}(E)=\frac{1}{2 \pi i}\int
    \limits _{\mathcal{C}_1}d\mathcal{E}\,G_{m_1 n_1}^{\ell_1(+)}(\sqrt{2\mathcal{E}})
    G_{m_2 n_2}^{\ell_2(+)}(\sqrt{2(E-\mathcal{E})})
\end{equation}
of two one-particle Green's function $G^{\ell(+)}$ matrix elements.
The latter is expressed in terms of two independent $J$-matrix
solutions \cite{Heller:1975,JM1}:
\begin{equation}\label{Gnm2}
    G^{\ell(+)}_{m n}(k)=-\frac{2}{k}S_{n_{<} \ell}(k)C_{n_{>}
    \ell}^{(+)}(k),
\end{equation}
\begin{equation}
 \begin{array}{c}\label{Ssol}
S_{n \ell}(k)= \frac{1}{2}\left[(n+1)_{(2 \ell+1)}
\right]^{1/2}\,(2\sin \xi)^{\ell+1}\,e^{-\pi
\beta/2}\,\omega^{-{i}\beta}\frac{\left|\Gamma(\ell+1+{i}\beta)
\right|} {(2\ell+1)!}\\[3mm]
\times (-\omega)^n\,{_2F_1} \left( -n, \ell+1+ {i} \beta; 2 \ell+2;
1-\omega^{-2} \right),
 \end{array}
\end{equation}
\begin{equation}\label{Cplus}
 \begin{array}{c}
    C_{n\ell}^{(+)}(k) = -\sqrt{n!(n+2\ell+1)}\frac{e^{\pi \beta/2}\omega^{i \beta}}{(2 \sin
    \xi)^{\ell}}\\
 \times \frac{\Gamma(\ell+1+i \beta)}{|\Gamma(\ell+1+i
 \beta)|}\frac{(-\omega)^{n+1}}{\Gamma(n+\ell+2+i \beta)}{_2F_1\left(-\ell+i \beta,
 n+1;n+\ell+2+i \beta; \omega^2  \right)},
\end{array}
\end{equation}
where
\begin{equation}\label{omegaz}
 \omega\equiv e^{i \xi}=\frac{b+i k}{b-i k}, \quad
 \sin \xi = \frac{2 b k }{b^2+k^2},
\end{equation}
$\beta=\frac{-2}{k}$ is the Sommerfeld parameter. Following the
method of \cite{Shakeshaft:2004}, we perform the integration in
(\ref{GMECI}) along contour $\mathcal{C}_1$ (see Fig.~\ref{fig1})
which is obtained by rotating the contour $\mathcal{C}$ by angle
$-\pi < \varphi < 0$ about the point $\frac{E}{2}$. This allows us
to avoid the singularities of $G^{\ell_1(+)}_{m_1 n_1}$.

Figs.~\ref{fig3} and \ref{fig4} show the behavior of $Q^{0 0(+)}_{0
0}$ (\ref{CQSS}) (the dashed lines) along the
$r_1=r_2=\rho/\sqrt{2}$ diagonal. For these calculations, we choose
$E=0.735$ a.u. and 25 for the upper limit of the sum. The scale
parameter $b$ was set to 1.6875 and the rotation angle $\varphi$ was
$-\frac{\pi}{3}$.

\subsection{Asymptotic Behavior}

The asymptotic form of the QS function $Q^{\ell(\pm)}_{n}$ can be
written as \cite{DelPunta:2014}
\begin{equation}
\label{Qasym}
  Q^{\ell(\pm)}_n(k, r)\mathop{\sim} \limits _{r \rightarrow \infty}
    -\frac{2}{k}\,S_{n \ell}(k)\,e^{\pm i\left(k r-\beta \ln(2 k r)-\frac{\pi
    \ell}{2}+\sigma_{\ell}(k) \right)},
\end{equation}
where
$$
e^{i\sigma_{\ell}(k)}=\frac{\Gamma(\ell+1+{i}\beta)}{\left|\Gamma(\ell+1+{i}\beta)\right|}.
$$

The asymptotic behavior of the CQS function (\ref{QC}) for $r_1
\rightarrow \infty$ and $r_2 \rightarrow \infty$ simultaneously (in
the constant ratio $\tan (\alpha) = r_2/r_1$, where $\alpha$ is the
hyperangle) is obtained by replacing $Q^{\ell_1(+)}_{n_1}$ and
$Q^{\ell_2(+)}_{n_2}$ by their asymptotic approximation
(\ref{Qasym}) and making use of the stationary phase method to
evaluate the resulting integral. The stationary point
$\mathcal{E}_0$ which satisfies the equation (see, e. g.,
\cite{Merkuriev_Faddeev})
\begin{equation}\label{SP}
    \frac{\partial}{\partial \mathcal{E}}\left(\sqrt{\mathcal{E}}r_1+ \sqrt{E-\mathcal{E}}r_2\right)=0.
\end{equation}
is $\mathcal{E}_0=\cos^2(\alpha)E$. Therefore, we finally obtain
\begin{equation}
\label{QxyA}
\begin{array}{c}
  Q_{n_1 n_2}^{\ell_1 \ell_2(+)}(E; r_1, r_2)
   \mathop{\sim} \limits _{\rho \rightarrow \infty}
    \frac{1}{E}\sqrt{\frac{2}{\pi}}(2E)^{3/4}e^{\frac{i
    \pi}{4}}S_{n_1 \ell_1}(p_1)S_{n_2 \ell_2}(p_2)\frac{1}{\sqrt{\rho}}\\
\times    \exp\left\{i\left[\sqrt{2E}\rho
    -\beta_1\ln(2p_1 r_1)-\beta_2\ln(2p_2 r_2)+\sigma_{\ell_1}(p_1)+\sigma_{\ell_2}(p_2)
    -\frac{\pi(\ell_1+\ell_2)}{2} \right] \right\},\\
 \end{array}
\end{equation}
where $\rho=\sqrt{r_1^2+r_2^2}$ is the hyper-radius,
$p_1=\cos(\alpha)\sqrt{2E}$, $p_2=\sin(\alpha)\sqrt{2E}$,
$\beta_{1,2}= \frac{-2}{p_{1,2}}$.

\subsection{Integral representation}

Rather than apply the expansion (\ref{CQSS}) to calculate the CQS
functions at large distances, it might be more convenient to employ
the contour integral (\ref{QC}) whose integrand is expressed in
terms of the integral \cite{DelPunta:2014}
\begin{equation}\label{QIR}
  \begin{array}{c}
    Q^{\ell(\pm)}_{n}(k, r) = -\left[(n+1)_{2\ell+1}
    \right]^{-\frac{1}{2}}\,(2 b r)^{\ell+1}\,e^{-b r}\frac{2}{(b \mp ik)}\,\int \limits
    _{0}^{1}d z \, (1-z)^{\ell \pm i \alpha}(1-\omega^{\pm 1}z)^{\ell \mp i \alpha}\\
    \times (1-z-\omega^{\pm
    1}z)^n \exp\left(z\left[b \pm i k \right]r \right)
    \,L_{n}^{2 \ell + 1}\left(\frac{(1-z)(1-\omega^{\pm 1}z)}{(1-z-\omega^{\pm
    1}z)}\,2 b r \right).\\
  \end{array}
\end{equation}

\begin{figure*}[ht]
\centerline{\psfig{figure=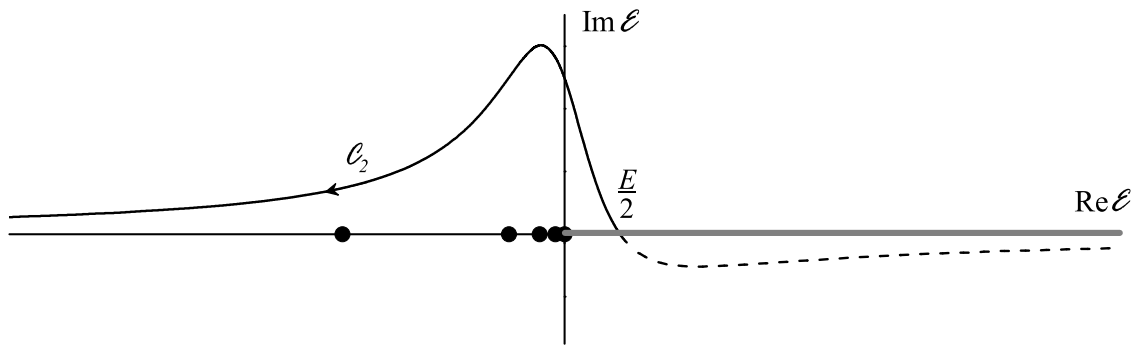,width=1\textwidth}}
\caption{The deformed contour $\mathcal{C}_{2}$ asymptotically
approaches the real energy axis.} \label{fig2}
\end{figure*}

\begin{figure*}[ht]
\centerline{\psfig{figure=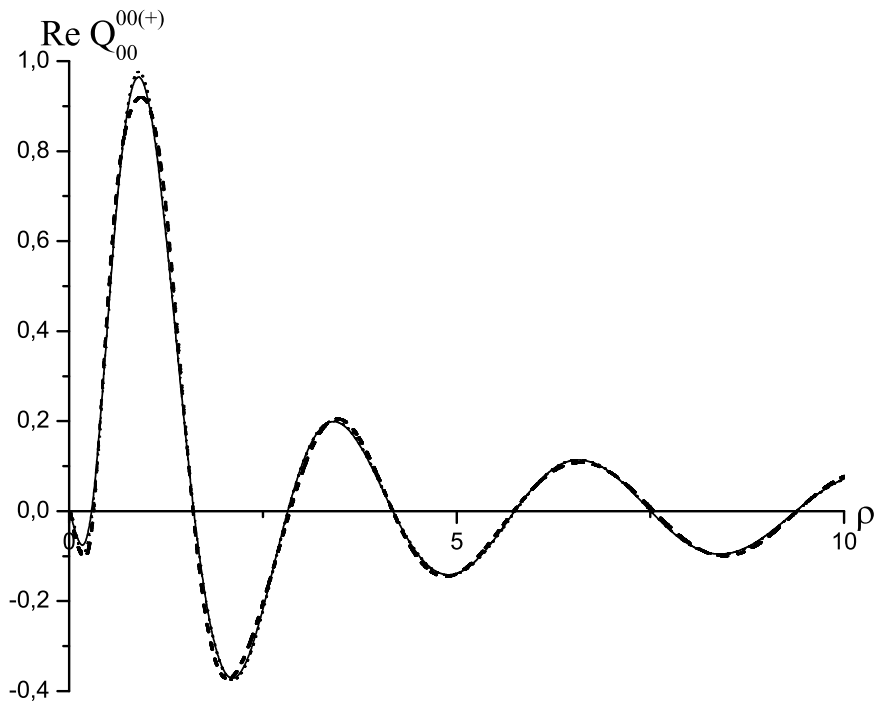,width=0.8\textwidth}}
\caption{The real part of the $s$-wave CQS function $Q_{00}^{00(+)}$
for $E=0.735$ and $b=1.6875$ along the diagonal
$r_1=r_2=\rho/\sqrt{2}$, approximated by (\ref{CQSS}) with the upper
limit $25$ (dashed line) and obtained by integrating (\ref{QC})
along the contour $\mathcal{C}_2$ (\ref{EC}) with $D=0.85$ (solid
line) and $D=15$ (dotted line) using the analytic continuation.}
\label{fig3}
\end{figure*}

\begin{figure*}[ht]
\centerline{\psfig{figure=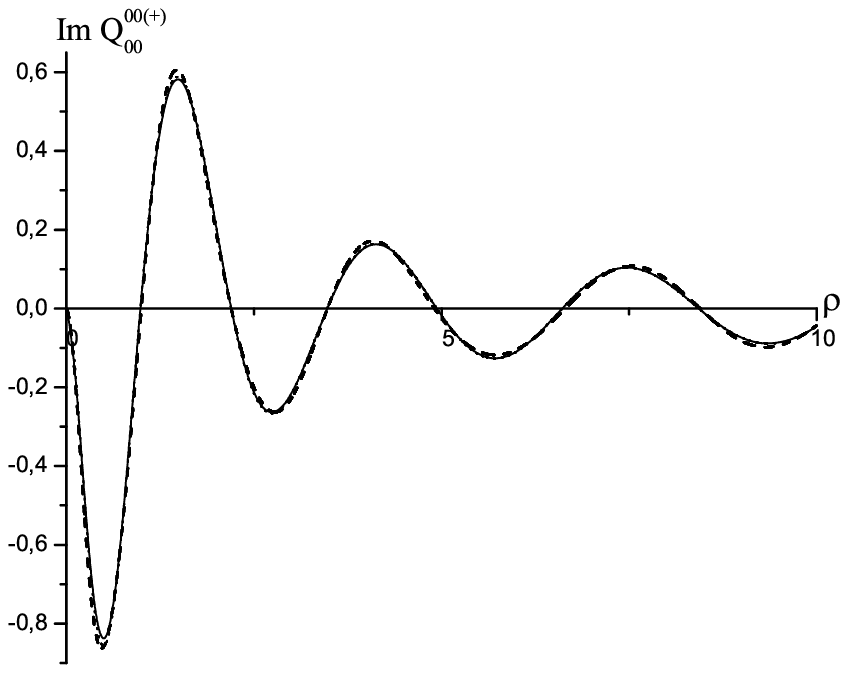,width=0.8\textwidth}}
\caption{The same as in Fig.~\ref {fig3} but for the imaginary
part.} \label{fig4}
\end{figure*}

Note that a part of the rotated straight-line contour
$\mathcal{C}_1$ indicated by a dashed line in Fig.~\ref{fig1} lies
in the unphysical energy sheet ($-2 \pi < \arg(\mathcal{E})<0$). In
turn, $Q^{\ell(+)}_n(\sqrt{2\mathcal{E}}, r)$ diverges exponentially
for large $\mbox{Im}(\mathcal{E})$ (see, e.g., (\ref{Qasym})) in the
lower half-plane. In order to ensure convergence of the integral
(\ref{QC}) we deform the contour $\mathcal{C}_1$ in such a way that
the resulting path $\mathcal{C}_2$, shown in Fig.~\ref{fig2},
asymptotically approaches the real axis. Specifically, the energy
$\mathcal{E}$ on the contour $\mathcal{C}_2$ is parametrized in the
form
\begin{equation}\label{EC}
    \mathcal{E}=t+i D \frac{(\frac{E}{2}-t)}{1+t^2},
\end{equation}
where $D$ is a positive constant and $t$ runs from $\infty$ to
$-\infty$. Generally there are points on the contour $\mathcal{C}_2$
at which $\mbox{Re}(\ell+i\beta)<-1$ and thus the integrand in
(\ref{QIR}) is singular at the endpoint $z=1$. To avoid this, we
apply the following procedure. Let $m$ be the minimum positive
integer number such that $-m< \mbox{Re}(\ell+i\beta)$. Then
integrating (\ref{QIR}) by part $m-1$ times and assuming the
integrated terms vanish at the limit $z=1$ we obtain the integral
which can be evaluated numerically. As a test of this analytic
continuation of $Q^{\ell(+)}_{n}(k, r)$ into the complex $k$-plane,
the CQS function $Q_{0 0}^{0 0(+)}$ has been calculated using the
integral (\ref{QC}) along the contour $\mathcal{C}_2$. We choose
$D=0.85$ for which $m=3$. The results for real and imaginary parts
are plotted in Fig.~\ref{fig3} and Fig.~\ref{fig4} by the solid
lines. We also plot the CQS function $Q_{0 0}^{0 0(+)}$ calculated
with $D=15$ for which $-1< \mbox{Re}(i\beta)$, i.e., $m=1$ (the
dashed lines). Agreement between these representations of CQS
functions and (\ref{CQSS}) argues for the suggested analytic
continuation.

\begin{figure*}[ht]
\centerline{\psfig{figure=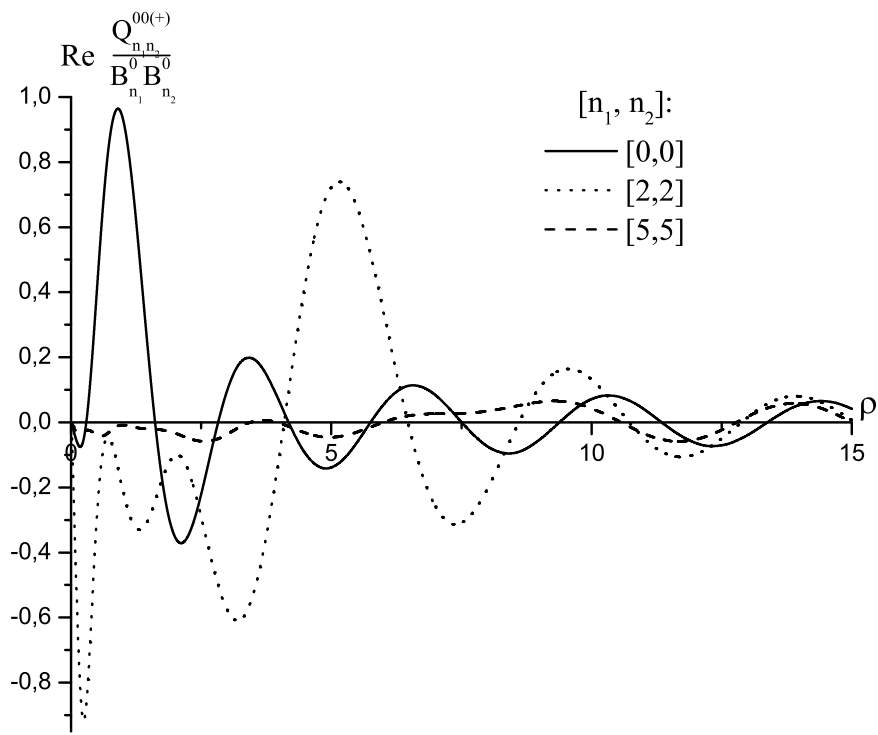,width=0.8\textwidth}}
\caption{Real parts for the first few $s$-wave CQS functions
$\frac{Q_{n_1 n_2}^{00(+)}\left(E; r_1,
r_2\right)}{B^0_{n_1}(p_1)B^0_{n_2}(p_2)}$ along the $r_1=r_2$
diagonal.} \label{fig5}
\end{figure*}

\begin{figure*}[ht]
\centerline{\psfig{figure=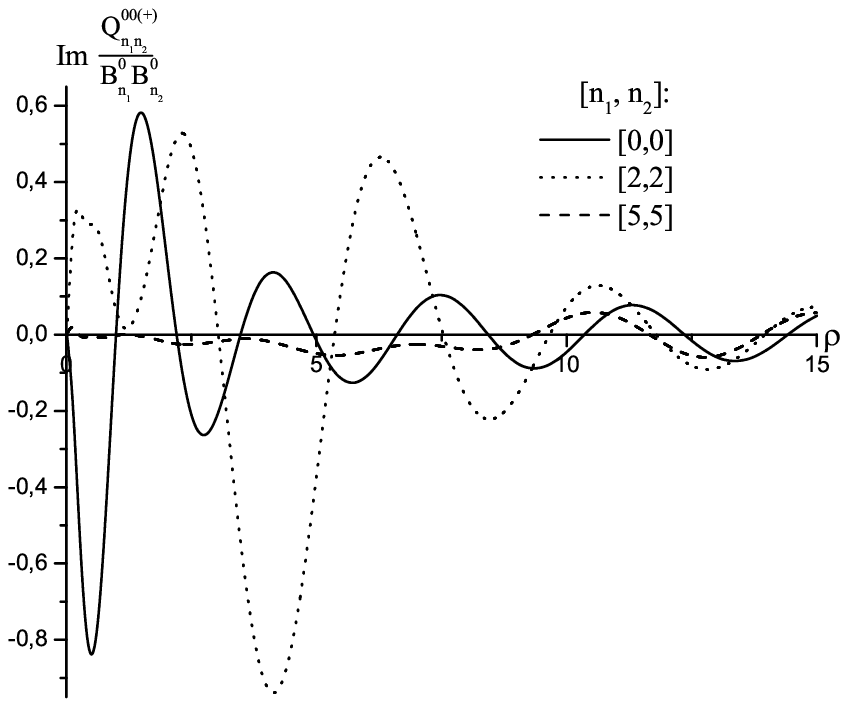,width=0.8\textwidth}}
\caption{The same as in Fig.~\ref{fig5} but for the imaginary
parts.} \label{fig6}
\end{figure*}

\begin{figure*}[ht]
\centerline{\psfig{figure=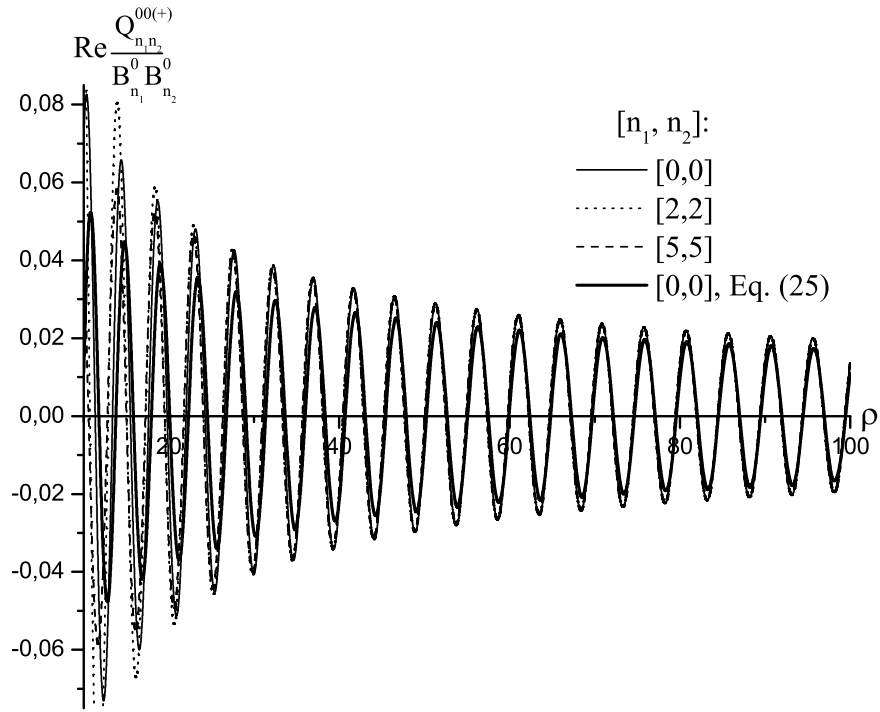,width=0.8\textwidth}}
\caption{Comparison of the real parts for the first few CQS
functions $\frac{Q_{n_1 n_2}^{00(+)}\left(E; r_1,
r_2\right)}{B^0_{n_1}(p_1)B^0_{n_2}(p_2)}$ and that of the
asymptotic limit (\ref{QxyA}) for $Q_{00}^{(00)(+)}$.} \label{fig7}
\end{figure*}

\begin{figure*}[ht]
\centerline{\psfig{figure=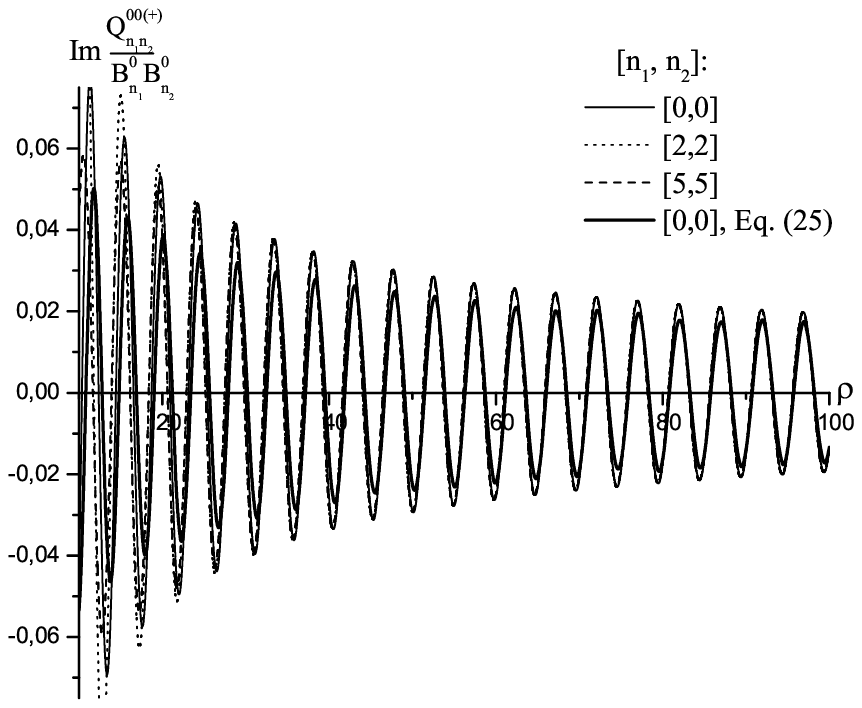,width=0.8\textwidth}}
\caption{The same as in Fig.~\ref{fig7} but for the imaginary
parts.} \label{fig8}
\end{figure*}

Note that the asymptotic behavior (\ref{QxyA}) of the two-particle
CQS functions depends upon the indices $n_1$ and $n_2$. It follows
from (\ref{Ssol}) that this dependence can be eliminated by dividing
(\ref{QC}) by $B^{\ell_1}_{n_1}(p_1)B^{\ell_2}_{n_2}(p_2)$, where
\begin{equation}\label{Bn}
    B^{\ell}_{n}(k)=[(n+1)_{2\ell+1}]^{1/2}(-\omega)^n\,{_2F_1} \left( -n,
\ell+1+ {i} \beta; 2 \ell+2; 1-\omega^{-2} \right).
\end{equation}
We present in Figs.~\ref{fig5} and \ref{fig6} a few $s$-wave CQS
functions $Q_{n_1 n_2}^{00(+)}/B^{0}_{n_1}(p_1)B^{0}_{n_2}(p_2)$ for
$\alpha=\frac{\pi}{4}$. These functions asymptotic behavior at large
distances is shown in Figs.~\ref{fig7} and \ref{fig8}. For
comparison, we also show the asymptotic approximation (\ref{QxyA})
for $Q_{0 0}^{00(+)}$.

Note that it follows from (\ref{QIR}) that on the left part of the
contour $\mathcal{C}_2$ where $k \sim i|k|$ and $|k| \rightarrow
\infty$ the function $Q^{\ell(+}_{n}$ for large $r$ behaves like
$e^{-br}$ rather than $e^{i k r}$. Thus, the greater is the scale
parameter $b$, the faster the CQS function (\ref{QC}) reaches its
asymptotic form (\ref{QxyA}).

\subsection{The solution asymptotic form}

We try to solve the equation (\ref{ME}) by an expansion into the
basis set of CQS functions (\ref{QSANG}) whose asymptotic behavior
in the region $\Omega_0$ is not correct since it misses out at least
the phase factor, corresponding to the Coulomb interelectronic
interaction (see, e. g., \cite{Rudge,Merkuriev_Faddeev}):
\begin{equation}\label{W3}
  W_3({\bf r}_1, {\bf r}_2)=-\frac{\rho}{\sqrt{2E}}\frac{1}{r_{12}}\ln \left(2\sqrt{2E}\rho\right).
\end{equation}
Inserting (\ref{QxyA}) into (\ref{FE}) we find the formal result for
the the asymptotic form of the solution of (\ref{ME}) at large
distances:
\begin{equation}\label{FA}
 \begin{array}{c}
 \Phi^{(+)}({\bf r}_1, {\bf r}_2)\approx
\frac{2}{E\sin(2\phi)}\sqrt{\frac{2}{\pi}}(2E)^{3/4}e^{\frac{i
\pi}{4}}\frac{\exp\left\{i\left[\sqrt{2E}\rho-\alpha_1\ln(2p_1
r_1)-\alpha_2\ln(2p_2 r_2)\right]\right\}} {\rho^{5/2}}\\
\times \sum \limits _{\ell_1 \ell_2 L}\mathcal{Y}^{\ell_1
\ell_2}_{LM}(\hat{\bf r}_1,\hat{\bf
r}_2)\exp\left\{i\left[\sigma_{\ell_1}(p_1)+\sigma_{\ell_2}(p_2)
    -\frac{\pi(\ell_1+\ell_2)}{2} \right]  \right\}\\
\times \sum \limits _{n_1, n_2=0}^{N-1} C_{n_1 n_2}^{L(\ell_1
\ell_2)}S_{n_1 \ell_1}(p_1)S_{n_2 \ell_2}(p_2).\\
 \end{array}
\end{equation}

Thus the feasibility of using the CQS functions (\ref{QSANG}) for
solving (\ref{ME}) is questionable. To improve the asymptotic
properties of the CQS function (and thereby solve the problem of
slow convergence of the expansion (\ref{FE})), it may be useful to
modify the CQS function by multiplying it by
\begin{equation}\label{EWl}
 e^{i \mathcal{W}_{\ell_1
\ell_2}\left(r_1,r_2\right)},
\end{equation}
where
\begin{equation}\label{Wl}
    \mathcal{W}_{\ell_1
\ell_2}\left(r_1,r_2\right)\mathop{\sim}\limits_{\rho \rightarrow
\infty}-\frac{\rho}{\sqrt{2E}}\,\mathcal{V}_{\ell_1
\ell_2}\left(r_1,r_2\right)\ln \left(2\sqrt{2E}\rho\right),
\end{equation}
\begin{equation}\label{Vl}
    \mathcal{V}_{\ell_1
\ell_2}\left(r_1,r_2\right)=\int d\hat{\bf r}_1 d\hat{\bf
r}_2\left[\mathcal{Y}^{\ell_1 \ell_2}_{LM}(\hat{\bf r}_1,\hat{\bf
r}_2)\right]^{\ast}\frac{1}{r_{12}}\,\mathcal{Y}^{\ell_1
\ell_2}_{LM}(\hat{\bf r}_1,\hat{\bf r}_2).
\end{equation}
Hence the modified function takes the form
\begin{equation}\label{QEWl}
    \widetilde{Q}^{\ell_1 \ell_2(+)}_{n_1,n_2}\left(E;\; r_1,r_2 \right) = e^{i \mathcal{W}_{\ell_1
\ell_2}\left(r_1,r_2\right)}Q^{\ell_1 \ell_2(+)}_{n_1,n_2}\left(E;\;
r_1,r_2 \right).
\end{equation}
Note, however, that such phase factors can not take into account the
off-diagonal elements
\begin{equation}\label{Vll_}
    \mathcal{V}_{\ell_1
\ell_2,\ell'_1 \ell'_2}\left(r_1,r_2\right)=\int d\hat{\bf r}_1
d\hat{\bf r}_2\left[\mathcal{Y}^{\ell_1 \ell_2}_{LM}(\hat{\bf
r}_1,\hat{\bf
r}_2)\right]^{\ast}\frac{1}{r_{12}}\,\mathcal{Y}^{\ell'_1
\ell'_2}_{LM}(\hat{\bf r}_1,\hat{\bf r}_2).
\end{equation}
Thus, it will probably be more convenient to choose the basis
functions of the form
\begin{equation}\label{QEW}
     e^{i W_3\left({\bf r}_1,{\bf r}_2\right)}\left|n_1 \ell_1 n_2 \ell_2; L M  \right>_Q.
\end{equation}
It is beyond the scope of this paper to discuss in detail the
modified basis functions and their applications. In this report we
restrict ourselves to a simple $s$-wave case.

\section{Solving of the driven equation}

In the  Temkin-Poet model the equation (\ref{ME}) reduces to
\cite{Gasaneo:2013}
\begin{equation}\label{SME}
    \left[E+\frac{1}{2}\frac{\partial^2}{\partial r_1^2}+\frac{1}{2}\frac{\partial^2}{\partial r_2^2}
     +\frac{2}{r_1}+\frac{2}{r_2}-\frac{1}{r_>}
     \right]\chi(r_1,r_2)=\mathcal{F}(r_1,r_2),
\end{equation}
where $\chi(r_1,r_2)=r_1r_2\phi^{(+)}(r_1,r_2)$, whereas the right
hand side is given by
\begin{equation}\label{SRHS}
\mathcal{F}(r_1,r_2)=
-\frac{1}{(2\pi)^3}\frac{4\pi}{q^2}\left[2-j_0(qr_1)-j_0(qr_2)\right]r_1r_2\frac{Z_e^3}{\pi}\,
e^{-Z_e(r_1+r_2)}.
\end{equation}
We set $E=0.735$, $q=0.24$ and $Z_e=2-5/16$.

To solve the equation we first consider the expansion
\begin{equation}\label{ChiN}
    \chi(r_1,r_2)=\sum \limits
    _{n_1,n_2=0}^{N-1}C_{n_1,n_2}Q_{n_1,n_2}^{(+)}(r_1,r_2).
\end{equation}
Our discussion is limited to $s$ waves, so that we omit the
angular-momentum labels $\ell_1$ and $\ell_2$. We choose $Z_e$ for
the scale parameter $b$ of the basis. Inserting (\ref{ChiN}) into
(\ref{SME}) gives
\begin{equation}\label{SME2}
    \sum \limits_{n_1,n_2=0}^{N-1}\left[\frac{\psi_{n_1}(r_1)\psi_{n_2}(r_2)}{r_1r_2}
    -\frac{1}{r_>}\,Q_{n_1,n_2}^{(+)}(r_1,r_2)
     \right]C_{n_1,n_2}=\mathcal{F}(r_1,r_2).
\end{equation}
Then, multiplying Eq. (\ref{SME2}) by
$\psi_{m_1}(r_1)\psi_{m_2}(r_2)$ and integrating, in view of the
orthogonality condition (\ref{Orth}), gives
\begin{equation}\label{SME3}
    \sum
    \limits_{n_1,n_2=0}^{N-1}\left[\delta_{m_1,n_1}\delta_{m_2,n_2}
    +\mathcal{L}_{m_1,m_2;n_1,n_2}\right]C_{n_1,n_2}=\mathcal{R}_{m_1,m_2}, \quad
    m_1,m_2\leq N-1,
\end{equation}
where
\begin{equation}\label{Rhs}
    \mathcal{R}_{m_1,m_2}=\int \limits_{0}^{\infty}\int
    \limits_{0}^{\infty}dr_1\,dr_2\psi_{m_1}(r_1)\psi_{m_2}(r_2)\mathcal{F}(r_1,r_2).
\end{equation}
Further, we approximate the matrix elements
\begin{equation}\label{L1}
    \mathcal{L}_{m_1,m_2;n_1,n_2} \equiv \int \limits_{0}^{\infty}\int
    \limits_{0}^{\infty}dr_1\,dr_2\psi_{m_1}(r_1)\psi_{m_2}(r_2)\left(-\frac{1}{r_>}
    \right)Q_{n_1,n_2}^{(+)}(r_1,r_2)
\end{equation}
by using the expansion of CQS function into the Laguerre basis
(\ref{CQSS}) and taking into account the basis completeness:
\begin{equation}\label{L2}
    \mathcal{L}_{m_1,m_2;n_1,n_2} =  -\sum \limits_{n'_1,n'_2=0}^{N-1}
    V_{m_1,m_2;n'_1,n'_2}G^{(+)}_{n'_1,n'_2;n_1,n_2},
\end{equation}
where $V_{m_1,m_2;n'_1,n'_2}$ are the matrix elements of the $e-e$
interaction:
\begin{equation}\label{U3}
 V_{m_1,m_2;n'_1,n'_2} = \int \limits_{0}^{\infty}\int
    \limits_{0}^{\infty}dr_1\,dr_2\psi_{m_1}(r_1)\psi_{m_2}(r_2)\frac{1}{r_>}
    \psi_{n'_1}(r_1)\psi_{n'_2}(r_2).
\end{equation}

Our aim is to study the convergence properties of the expansion
(\ref{ChiN}) (in conjunction with the approximation (\ref{L2})) as
$N$ is increased. The real and imaginary parts of the solution
$\chi(r_1,r_2)\rho^{1/2}2/\sin 2\alpha=\phi^{(+)}\rho^{5/2}$ along
the $r_1=r_2=\rho/\sqrt{2}$ diagonal are shown in Figs. 9 and 10.
Figures 11 and 12 show the results for the asymptotic approximation
(\ref{FA}) (multiplied by $\rho^{5/2}$) to the solution. From the
plots we see that applying the expansion (\ref{ChiN}) yields a
solution with divergent phase as a function of $N$, whereas the
magnitude
\begin{equation}\label{A1}
    A \equiv
    \mathop{\lim}\limits_{\rho\rightarrow \infty}
    \left|\phi^{(+)}(r_1,r_2)\rho^{5/2}\right|
\end{equation}
seems to converge. Actually, from the asymptotic form (\ref{FA}) of
the solution, it follows that
\begin{equation}\label{AN}
    A_N = \frac{2 (2E)^{3/4}}{E
    \sin2\alpha}\sqrt{\frac{2}{\pi}}\frac{1}{4\pi}\left|\sum
    \limits_{n_1,n_2=0}^{N-1}C_{n_1,n_2}S_{n_1}(p_1)S_{n_2}(p_2)
    \right|.
\end{equation}
In the case $\alpha=\frac{\pi}{4}$, $p_1=p_2=\sqrt{E}$ we have
obtained $A_{16}=1.505 \times 10^{-4}$, $A_{21}=1.507 \times
10^{-4}$ and $A_{26}=1.400 \times 10^{-4}$.

It might be expected that the convergence could be achieved using
the modified CQS functions:
\begin{equation}\label{MCQS}
  \widetilde{Q}_{n_1,n_2}^{(+)}(r_1, r_2) \equiv e^{i\mathcal{W}(r_1,r_2)}Q_{n_1,n_2}^{(+)}(r_1,
  r_2),
\end{equation}
\begin{equation}\label{SW3}
  \mathcal{W}(r_1, r_2)=-\frac{\rho}{\sqrt{2E}}\frac{1}{(1+r_>)}\ln
  \left(2\sqrt{2E}(1+\rho)\right).
\end{equation}
Then substituting the expansion
\begin{equation}\label{ChiNt}
   \widetilde{\chi}(r_1,r_2)=\sum \limits
    _{n_1,n_2=0}^{N-1}\widetilde{C}_{n_1,n_2}\widetilde{Q}_{n_1,n_2}^{(+)}(r_1,r_2).
\end{equation}
into (\ref{SME}) gives
\begin{equation}\label{MEP}
  \sum \limits _{n_1,n_2=0}^{N-1}\left[\frac{\psi_{n_1}(r_1)\psi_{n_2}(r_2)}{r_1r_2}-\hat{U}Q_{n_1,n_2}^{(+)}
  (E;\;r_1,r_2) \right]\widetilde{C}_{n_1,n_2}=e^{-i \mathcal{W}(r_1,r_2)}\mathcal{F}(r_1,r_2),
\end{equation}
where the operator $\hat{U}$ is defined by
\begin{equation}\label{OPV}
 \begin{array}{c}
  \hat{U}=\frac{1}{r_>}+\frac{1}{2}\left(\frac{\partial \mathcal{W}}{\partial r_1} \right)^2+\frac{1}{2}
  \left(\frac{\partial \mathcal{W}}{\partial r_2} \right)^2-\frac{i}{2}
  \left(\frac{\partial^2 \mathcal{W}}{\partial r_1^2}+\frac{\partial^2 \mathcal{W}}{\partial r_2^2} \right)\\
  -i\left[\frac{\partial \mathcal{W}}{\partial r_1}\frac{\partial}{\partial r_1}
  +\frac{\partial \mathcal{W}}{\partial r_2}\frac{\partial}{\partial r_2} \right].\\
 \end{array}
 \end{equation}
For large $\rho$ one finds from (\ref{QxyA}) that
\begin{equation}\label{Diff}
    \frac{\partial}{\partial r_{1,2}}\,Q_{n_1,n_2}^{(+)}(E;\;r_1,r_2)
    \sim
    i\sqrt{2E}\,\frac{r_{1,2}}{\rho}\,Q_{n_1,n_2}^{(+)}(E;\;r_1,r_2),
\end{equation}
and therefore the action of $\hat{U}$ on $Q_{n_1,n_2}^{(+)}$ in the
asymptotic region is reduced to multiplication by the `effective
potential'
\begin{equation}\label{Ueff}
  U^{eff}_{n_1,n_2}(r_1,r_2) \equiv\frac{\hat{U}Q_{n_1,n_2}^{(+)}(E;\;r_1,r_2)}{Q_{n_1,n_2}^{(+)}(E;\;r_1,r_2)},
\end{equation}
such that
\begin{equation}\label{UeffA}
  U^{eff}_{n_1,n_2}(r_1,r_2) \mathop{\sim}\limits_{\rho \rightarrow \infty}\left(\frac{\ln \left(2\sqrt{2E}
  \rho\right)}
  {\sqrt{2E}\rho}\right)^2.
\end{equation}
Figure \ref{fig13} shows plots of the $U^{eff}_{00}$ and
$U^{eff}_{55}$ on the diagonal $r_1=r_2$. For comparison we have
also plotted the $e-e$ potential $\frac{1}{r_>}$.

A matrix equation for the coefficients $\widetilde{C}_{n_1,n_2}$ is
obtained by multiplying (\ref{MEP}) from the left by
$\psi_{m_1}(r_1)\psi_{m_2}(r_2)$ and integrating over both
coordinates:
\begin{equation}\label{SME4}
    \sum
    \limits_{n_1,n_2=0}^{N-1}\left[\delta_{m_1,n_1}\delta_{m_2,n_2}
    +\widetilde{\mathcal{L}}_{m_1,m_2;n_1,n_2}\right]\widetilde{C}_{n_1,n_2}=\widetilde{\mathcal{R}}_{m_1,m_2}, \quad
    m_1,m_2\leq N-1,
\end{equation}
where
\begin{equation}\label{Rhs2}
    \widetilde{\mathcal{R}}_{m_1,m_2}=\int \limits_{0}^{\infty}\int
    \limits_{0}^{\infty}dr_1\,dr_2\psi_{m_1}(r_1)\psi_{m_2}(r_2)e^{-i\mathcal{W}(r_1,r_2)}\mathcal{F}(r_1,r_2),
\end{equation}
\begin{equation}\label{L2t}
    \widetilde{\mathcal{L}}_{m_1,m_2;n_1,n_2} =  -\sum \limits_{n'_1,n'_2=0}^{N-1}
    U_{m_1,m_2;n'_1,n'_2}G^{(+)}_{n'_1,n'_2;n_1,n_2},
\end{equation}
\begin{equation}\label{U4}
 U_{m_1,m_2;n'_1,n'_2} = \int \limits_{0}^{\infty}\int
    \limits_{0}^{\infty}dr_1\,dr_2\psi_{m_1}(r_1)\psi_{m_2}(r_2)\hat{U}
    \psi_{n'_1}(r_1)\psi_{n'_2}(r_2).
\end{equation}

Figures \ref{fig14} and \ref{fig15} show the solution
$\widetilde{\chi}(r_1,r_2)\rho^{1/2}2/\sin
2\alpha=\widetilde{\phi}^{(+)}\rho^{5/2}$ along the diagonal
$r_1=r_2$. In turn, figures \ref{fig16} and \ref{fig17} present the
real and imaginary components of the corresponding asymptotic form
\begin{equation}\label{Phit}
\begin{array}{c}
\widetilde{\phi}^{(+)}\rho^{5/2} \mathop{\sim} \limits _{\rho
\rightarrow \infty}\frac{2 (2E)^{3/4}}{E
    \sin2\alpha}\sqrt{\frac{2}{\pi}}\frac{1}{4\pi}\sum
 \limits_{n_1,n_2=0}^{N-1}\widetilde{C}_{n_1,n_2}S_{n_1}(p_1)S_{n_2}(p_2)
 \exp\left(-i\frac{\rho}{\sqrt{2E}}\frac{1}{r_>}\ln
  (2\sqrt{2E}\rho) \right)\\
\times  \exp\left\{i\left[\sqrt{2E}\rho-\alpha_1\ln(2p_1
r_1)-\alpha_2\ln(2p_2 r_2)+\sigma_0(p_1)+\sigma_0(p_2)+\frac{\pi}{4}
\right ]\right\}.\\
\end{array}
\end{equation}
From the plots one can see that the convergence can be achieved by
introducing a phase factor corresponding to the interelectronic
potential $1/r_{12}$. Furthermore, such a modification of the
asymptotic form of the basis functions could result in appreciable
change in the solution magnitude
\begin{equation}\label{At}
    \widetilde{A} \equiv
    \mathop{\lim}\limits_{\rho\rightarrow \infty}
    \left|\widetilde{\phi}^{(+)}(r_1,r_2)\rho^{5/2}\right|.
\end{equation}
For comparison, in this case we have obtained
$\widetilde{A}_{16}=7.346 \times 10^{-4}$, $\widetilde{A}_{21}=7.396
\times 10^{-4}$ and $\widetilde{A}_{26}=7.593 \times 10^{-4}$.

\section{Summery}
Two-particle basis functions --- labelled CQS --- are proposed. By
analogy with the Green's function of two non-interacting hydrogenic
atomic systems, they are expressed as a convolution integral of two
one-particle QS functions. We suggest an analytic continuation of
the QS functions into the entire complex $k$-plane in order to
perform the numerical contour integration. The asymptotic limit of
the CQS basis functions in the region $\Omega_0$ is expressed in
closed form as a six-dimensional outgoing spherical wave.

We study the application of the expansion into the CQS functions to
the solution of the inhomogeneous equation describing the double
ionization of helium by high-energy electron impact in the framework
of the Temkin-Poet model. Note that in constructing the basis
functions we do not take into account the interelectronic
interaction, and therefore the asymptotic behavior of the basis
functions is not correct. Hence, the driven equation whose solution
is expanded in terms of these basis functions is noncompact (due to
the Coulomb potential $\frac{1}{r_{12}}$ in the left hand side).
Thus the applicability of this approach is questionable. We show
that the problem of slow convergence (or even perhaps lack of
convergence) of the expansion can be solved by using the modified
CQS functions equipped with the phase factor corresponding to the
potential $\frac{1}{r_{12}}$. Moreover, the solutions which satisfy
the different boundary conditions differ appreciably in magnitude.
These results suggest that suitable basis functions can be obtained
by expanding products of the missing phase factor and CQS functions
in series of bispherical harmonics.

\section*{Acknowledgments}
Authors are very grateful to Dr. Gustavo Gasaneo for discussions and
his permanent interest to this topics. We are thankful to the
Computer Center, Far Eastern Branch of the Russian Academy of
Science (Khabarovsk, Russia) and the Computer Center, Universit\'{e}
de Lorraine (Metz, France) for generous rendering of computer
resources to our disposal.

\newpage

\newpage

\newpage

\newpage

\newpage

\newpage

\newpage

\newpage

\newpage
\begin{figure*}[ht]
\centerline{\psfig{figure=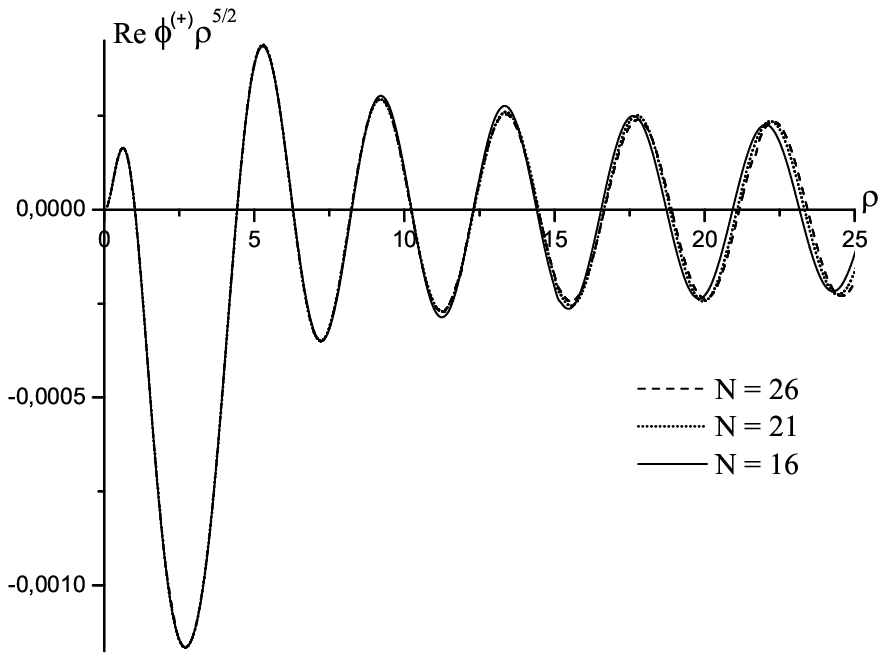,width=0.8\textwidth}}
\caption{The real components of the solutions $\phi^{(+)}\rho^{5/2}$
for different basis sizes along the diagonal $r_1=r_2$.}
\label{fig9}
\end{figure*}

\newpage
\begin{figure*}[ht]
\centerline{\psfig{figure=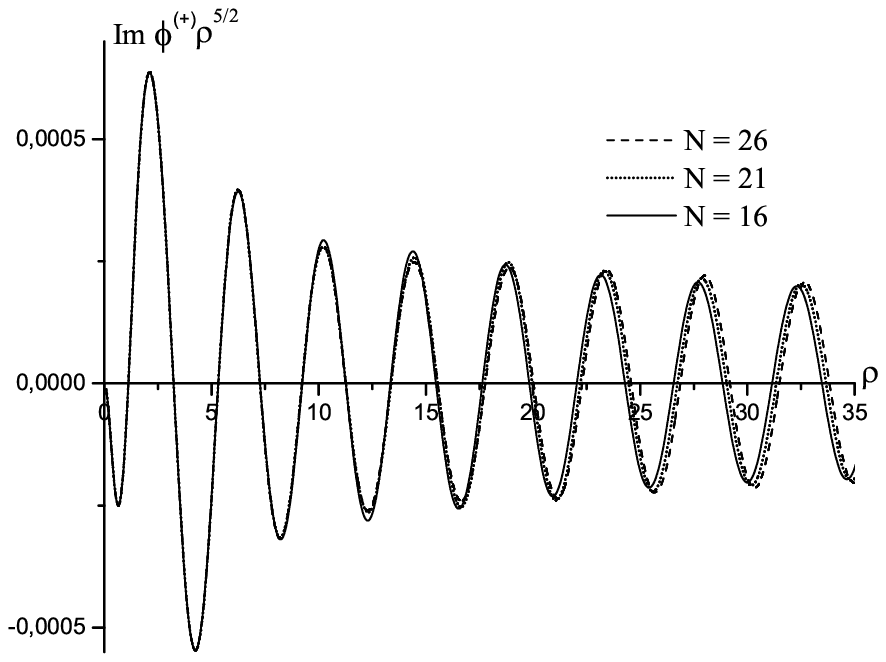,width=0.8\textwidth}}
\caption{The same as in Fig.~\ref{fig9} but for the imaginary
parts.} \label{fig10}
\end{figure*}

\newpage
\begin{figure*}[ht]
\centerline{\psfig{figure=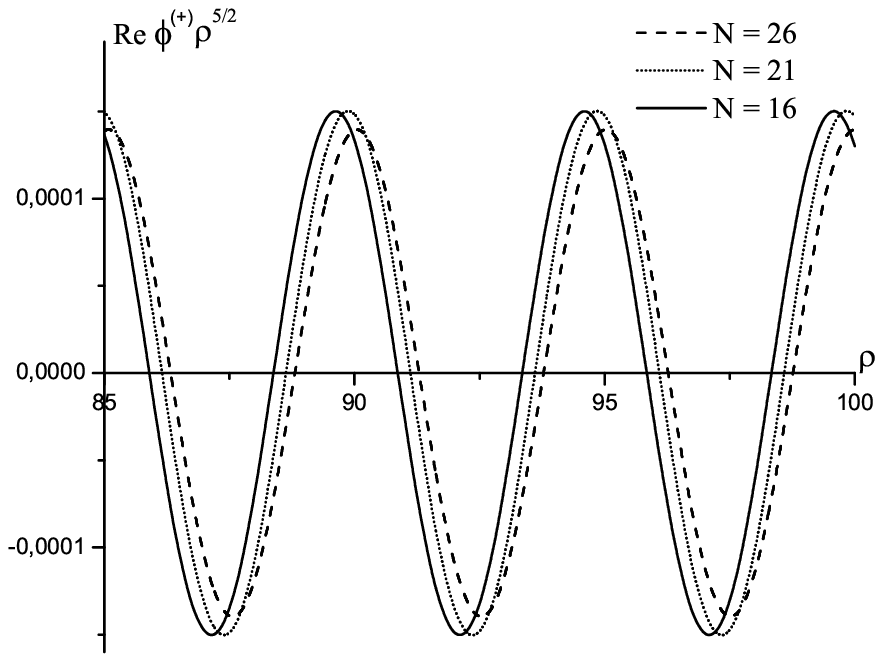,width=0.8\textwidth}}
\caption{Real parts of the asymptotic form for the solutions
$\phi^{(+)}\rho^{5/2}$ for different basis sizes along the diagonal
$r_1=r_2$.} \label{fig11}
\end{figure*}

\newpage
\begin{figure*}[ht]
\centerline{\psfig{figure=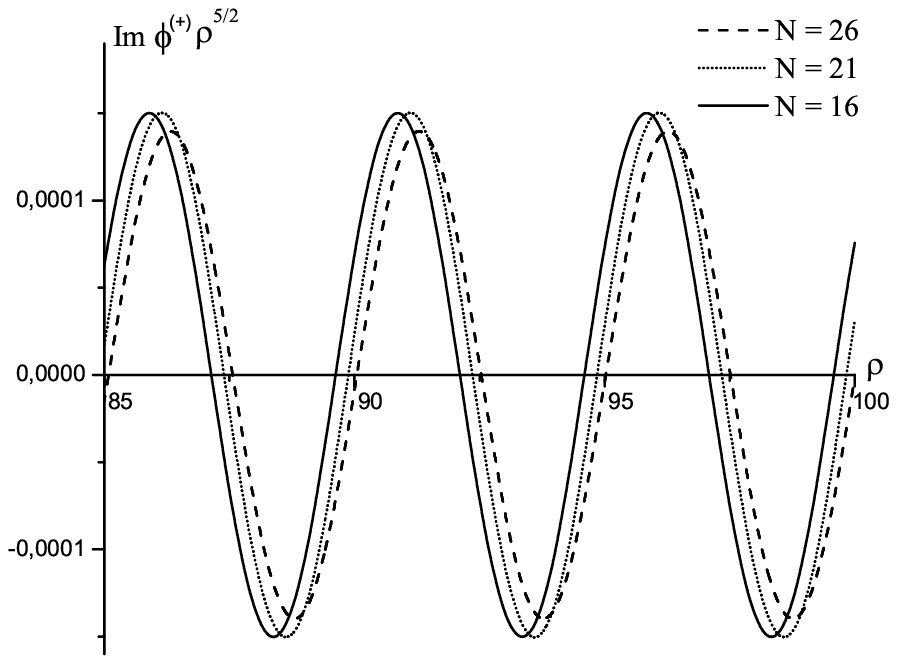,width=0.8\textwidth}}
\caption{The same as in Fig.~\ref{fig11} but for the imaginary
parts.} \label{fig12}
\end{figure*}

\newpage
\begin{figure*}[ht]
\centerline{\psfig{figure=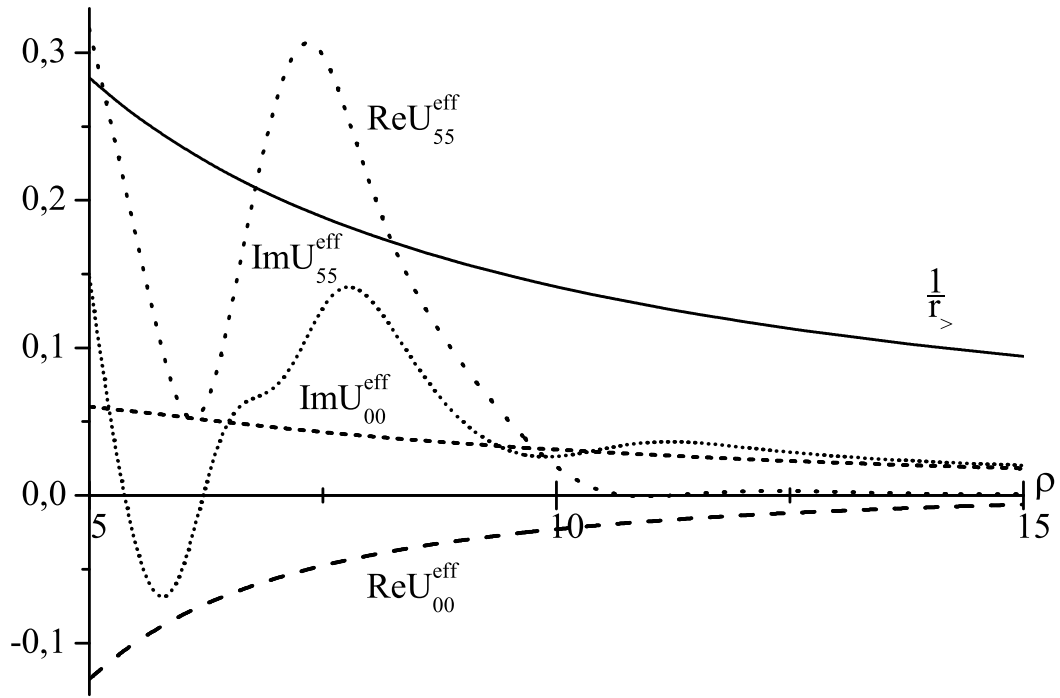,width=1\textwidth}} \caption{Real
and imaginary parts of the `effective potentials' $U_{00}^{eff}$ and
$U_{55}^{eff}$ (\ref{Ueff}).} \label{fig13}
\end{figure*}

\newpage
\begin{figure*}[ht]
\centerline{\psfig{figure=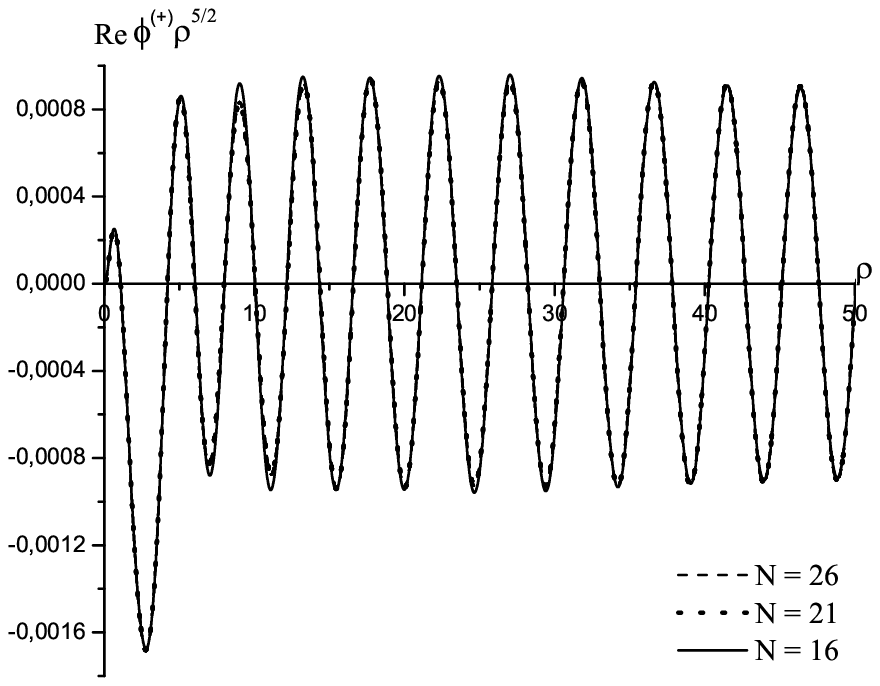,width=0.8\textwidth}}
\caption{The real components of the solutions
$\widetilde{\phi}^{(+)}\rho^{5/2}$ for different basis sizes along
the diagonal $r_1=r_2$.} \label{fig14}
\end{figure*}

\newpage
\begin{figure*}[ht]
\centerline{\psfig{figure=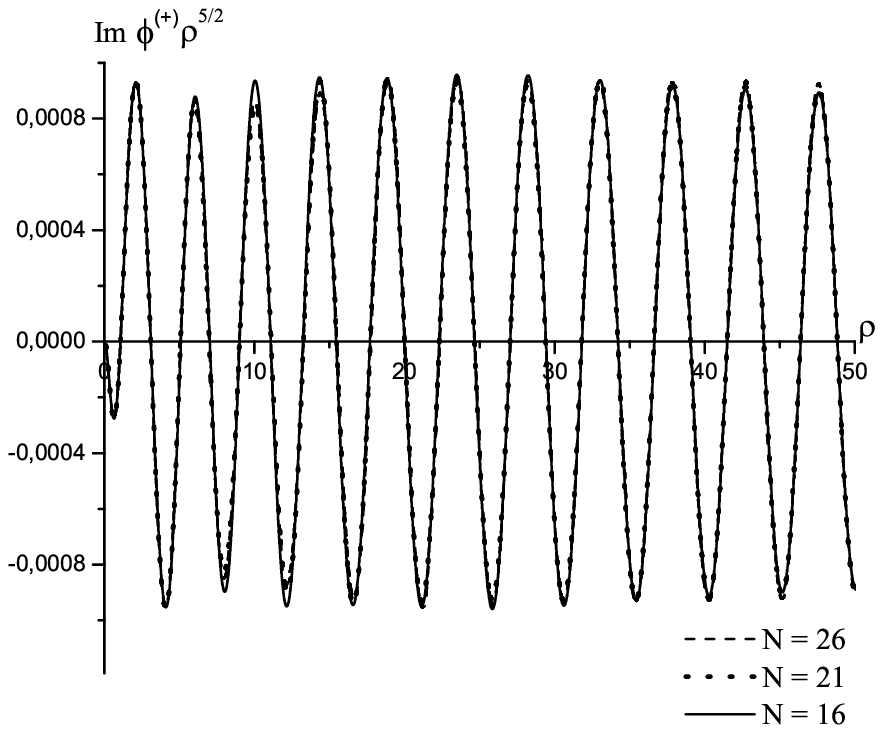,width=0.8\textwidth}}
\caption{The same as in Fig.~\ref{fig14} but for the imaginary
parts.} \label{fig15}
\end{figure*}

\newpage
\begin{figure*}[ht]
\centerline{\psfig{figure=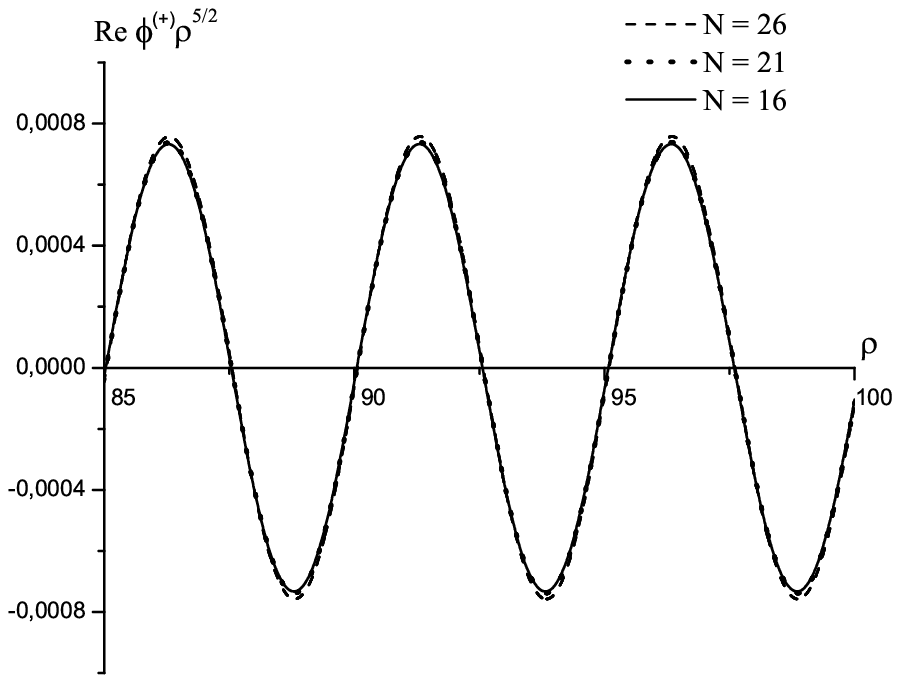,width=0.8\textwidth}}
\caption{Real parts of the asymptotic form for the solutions
$\widetilde{\phi}^{(+)}\rho^{5/2}$ (\ref{Phit}) for different basis
sizes along the diagonal $r_1=r_2$.} \label{fig16}
\end{figure*}

\newpage
\begin{figure*}[ht]
\centerline{\psfig{figure=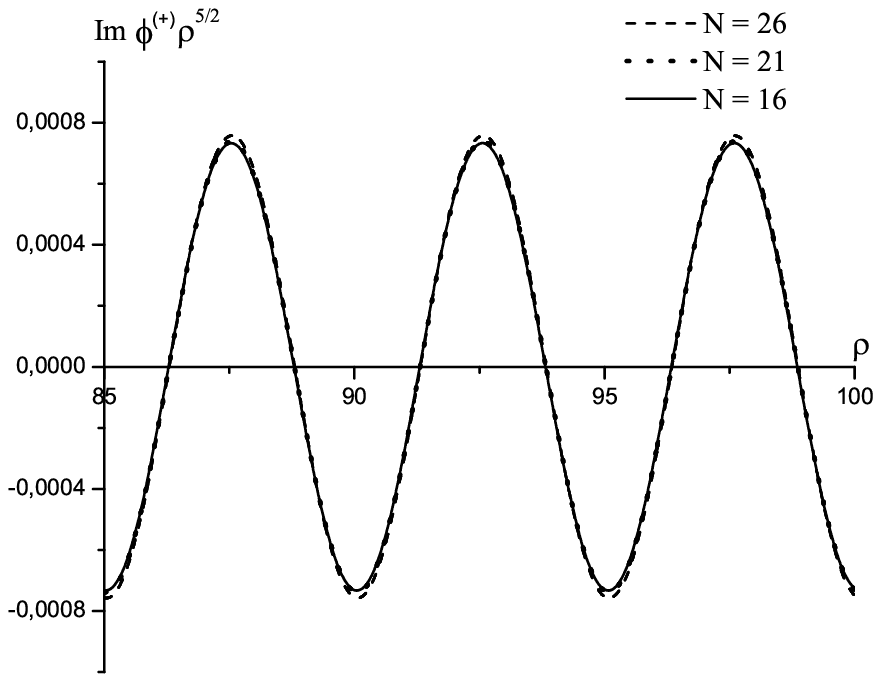,width=0.8\textwidth}}
\caption{The same as in Fig.~\ref{fig16} but for the imaginary
parts.} \label{fig17}
\end{figure*}

\end{document}